\documentclass[%
 reprint,
 amsmath,amssymb,
 aps,
]{revtex4-1}

\usepackage{graphicx}
\usepackage{dcolumn}
\usepackage{bm}

\usepackage{todonotes} 

\begin{document}

\preprint{APS/123-QED}

\title{Coupled nonlinear delay systems as deep convolutional neural networks}

\author{Bogdan Penkovsky}
 \altaffiliation[now at]{ Centre de Nanosciences et de Nanotechnologies, Univ. Paris-Sud, CNRS, Universit\'{e} Paris-Saclay, 91405 Orsay, France.}
\affiliation{FEMTO-ST/Optics Dept., UMR CNRS 6174, Univ. Bourgogne Franche-Comt\'{e}, 15B avenue des Montboucons, 25030 Besan\c{c}on Cedex, France}

\author{Xavier Porte}
\affiliation{FEMTO-ST/Optics Dept., UMR CNRS 6174, Univ. Bourgogne Franche-Comt\'{e}, 15B avenue des Montboucons, 25030 Besan\c{c}on Cedex, France}

\author{Maxime Jacquot}
\affiliation{FEMTO-ST/Optics Dept., UMR CNRS 6174, Univ. Bourgogne Franche-Comt\'{e}, 15B avenue des Montboucons, 25030 Besan\c{c}on Cedex, France}

\author{Laurent Larger}
\affiliation{FEMTO-ST/Optics Dept., UMR CNRS 6174, Univ. Bourgogne Franche-Comt\'{e}, 15B avenue des Montboucons, 25030 Besan\c{c}on Cedex, France}

\author{Daniel Brunner}
\affiliation{FEMTO-ST/Optics Dept., UMR CNRS 6174, Univ. Bourgogne Franche-Comt\'{e}, 15B avenue des Montboucons, 25030 Besan\c{c}on Cedex, France}

\date{\today}

\begin{abstract}

Neural networks are transforming the field of computer algorithms, yet their emulation on current computing substrates is highly inefficient.
 Reservoir computing was successfully implemented on a large variety of substrates and gave new insight in overcoming this implementation bottleneck.
 Despite its success, the approach lags behind the state of the art in deep learning.
 We therefore extend time-delay reservoirs to deep networks and demonstrate that these conceptually correspond to deep convolutional neural networks.
 Convolution is intrinsically realized on a substrate level by generic drive-response properties of dynamical systems.
 The resulting novelty is avoiding vector-matrix products between layers, which cause low efficiency in today's substrates.
 Compared to singleton time-delay reservoirs, our deep network achieves accuracy improvements by at least an order of magnitude in Mackey-Glass and Lorenz timeseries prediction.

\end{abstract}

\pacs{Valid PACS appear here}
\maketitle


Neural networks have emerged as the current disruptive computational concept.
 When cascading multiple network layers, these systems set the benchmark in multiple challenging tasks \cite{LeCun2015}.
 In such \textit{deep} neural networks, layers are dedicated to highlight specific aspects of the input-information, and previous layers commonly serve as input of consecutive layers.
 Such a hierarchical arrangement is crucial for boosting the computational performance.
 In deep convolutional neural networks (CNN), layers convolute their input with spatial filters.
 By increasing filter width and step size, deeper layers focus on more general features, while local features are highlighted in earlier layers \cite{Krizhevsky2012ImageNetNetworks}.

In the wake of deep neural networks' success, it was realized that their emulation on Turing / von Neumann machines is highly inefficient.
 This stimulated strong interest in the realization of neural networks in physical substrates whose architecture submit to the networks' topology.
 Particularly photonic systems, which offer key advantages for parallelization, are considered a promising future alternative.
 However, directly mapping the complex topology of a deep neural network onto a hardware substrates presents a significant challenge.
 Of essential importance are therefore concepts which strike a balance between architectural complexity and hardware implementation simplicity.

\begin{figure}[t]
\includegraphics[width=0.45\textwidth]{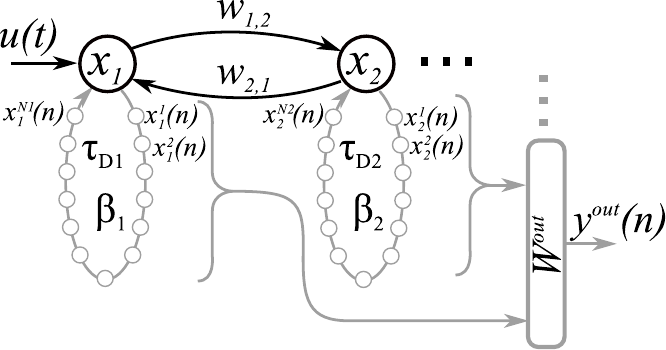}
    \caption{\label{fig:Scheme}
    Schematic of cascaded nonlinear oscillators acting as deep network, here consisting of two layers.
     Two coupled nonlinear delay systems $x_{1}(t)$ with states $x_{2}(t)$ implement individual time delay reservoir layers.
     Information is injected into the first system, and nonlinear nodes are coupled instantaneously according to weights $w_{1,2}$ and $w_{2,1}$.
     The readout-layer has access to all layers.}
\vspace{-0.7cm}     
\end{figure}

Among the various neural network architectures, reservoir computers \cite{Jaeger2004} have emerged as especially interesting theoretical model-systems \cite{Lu2018,Inubushi2017,Marzen2017} and promising candidates for hardware implementations.
 A reservoir computer is a complex recurrent neural network and conceptually corresponds to a high-dimensional nonlinear dynamical system.
 Training is restricted to the connections between the reservoir and its output, and hence the nonlinear dynamical system's topology remains constant.
 This strongly assists implementations in physical substrates, resulting in a large number of realizations in nonlinear photonic \cite{VanderSande2017} and other physical systems \cite{Tanaka2019RecentReview}.
 Yet, precisely this simplicity raised fundamental concerns regarding deep reservoirs.
 Recently it was found that, comparable to deep convolutional networks, a continuous change of \textit{spatial}-frequency in the response of consecutive  layers appears beneficial \cite{Gallicchio2016,Gallicchio2017}.
 The workhorse of the field have been nonlinear delay systems implementing time delay reservoirs (TDRs) \cite{VanderSande2017,Larger2017, Brunner2018}.
 These offer a compromise between good computing performance and exceptional ease of hardware implementation and serve as model-systems for more complex hardware substrates \cite{Shen2016, Bueno2018, Lin2018}.

We report on a deep reservoir scheme comprising hierarchically coupled nonlinear delay oscillators exhibiting dynamics on multiple timescales.
 Crucially, coupling between different layers is constant and training remains limited to the readout weights, in contrast to a proposed deep hardware TDR \cite{Nakajima2018}.
 This is an essential simplification as it adheres to the conceptual simplicity motive, which strongly fosters hardware implementation.
 We find that cascading significantly and qualitatively improves computational performance when compared to a single layer reservoir of identical size.
 Crucially, our architectural simplicity curbs the challenges particular to physically implementing complex and large networks.

In Fig. \ref{fig:Scheme}, we schematically illustrate our deep TDR concept.
 Dynamics are governed by the following set of equations:
\begin{align}
    &\tau_{i} \dot{x}_{i}(t) = -x_{i}(t) - \delta_{i}y_{i}(t) + \beta_{i}\sin^{2}[ d_{i}(t) + b_{i} ] \label{eq:DDIE} \\
    &\dot{y}_{i}(t) = x_{i}(t) \label{eq:Int} \\ 
    &d_{i}(t) = x_{i}(t-\tau_{Di}) + \sum_{p=\pm 1} w_{i+p,i}x_{i+p}(t) + \rho_{i} u(t) \label{eq:Arg} \\
    &\rho_{i\neq 1} = 0, \delta_{1} = 0. \label{eq:input}
\end{align}
 The state of the delay-coupled node in layer $i\in \{ 1, \cdots, I \}$ is given by $x_{i}(t)$, and we use the $\sin^{2}$-nonlinearity often employed in photonic TDRs \cite{Larger2012,Paquot2012}.
 Due to inertia, dynamics generally experience low-pass (LP) filtering according to a fast time constant $\tau_{i}$, which can be extended to band-pass (BP) filtering when a slow time constant $\delta_{i}$ is added \cite{Larger2013}.
 Each layer's nonlinearity is weighed by bifurcation parameter $\beta_{i}$, and the nonlinearity's argument contains constant bias $b_{i}$ and a time-dependent drive $d_{i}(t)$, see Eq. \eqref{eq:Arg}.
 Drive $d_{i}(t)$ features self-feedback delayed by $\tau_{Di}$ and potentially bidirectional coupling to adjacent layers according to coefficients $w_{i\pm 1, i}$.
 Only the first layer is coupled to $u(t)$, see Eq. \eqref{eq:input}.
 External drive $u(t)$ encodes the information to be processed $s(t)$ according to the temporal masking procedure which implements a linear matrix multiplication \cite{Appeltant2011, Brunner2018}.
 We have employed a de-synchronized information injection procedure in which each value of $s(t)$ is kept for an input-masking length of $0.8\cdot\tau_{Di}$.

\begin{figure}[t]
\includegraphics[width=0.5\textwidth]{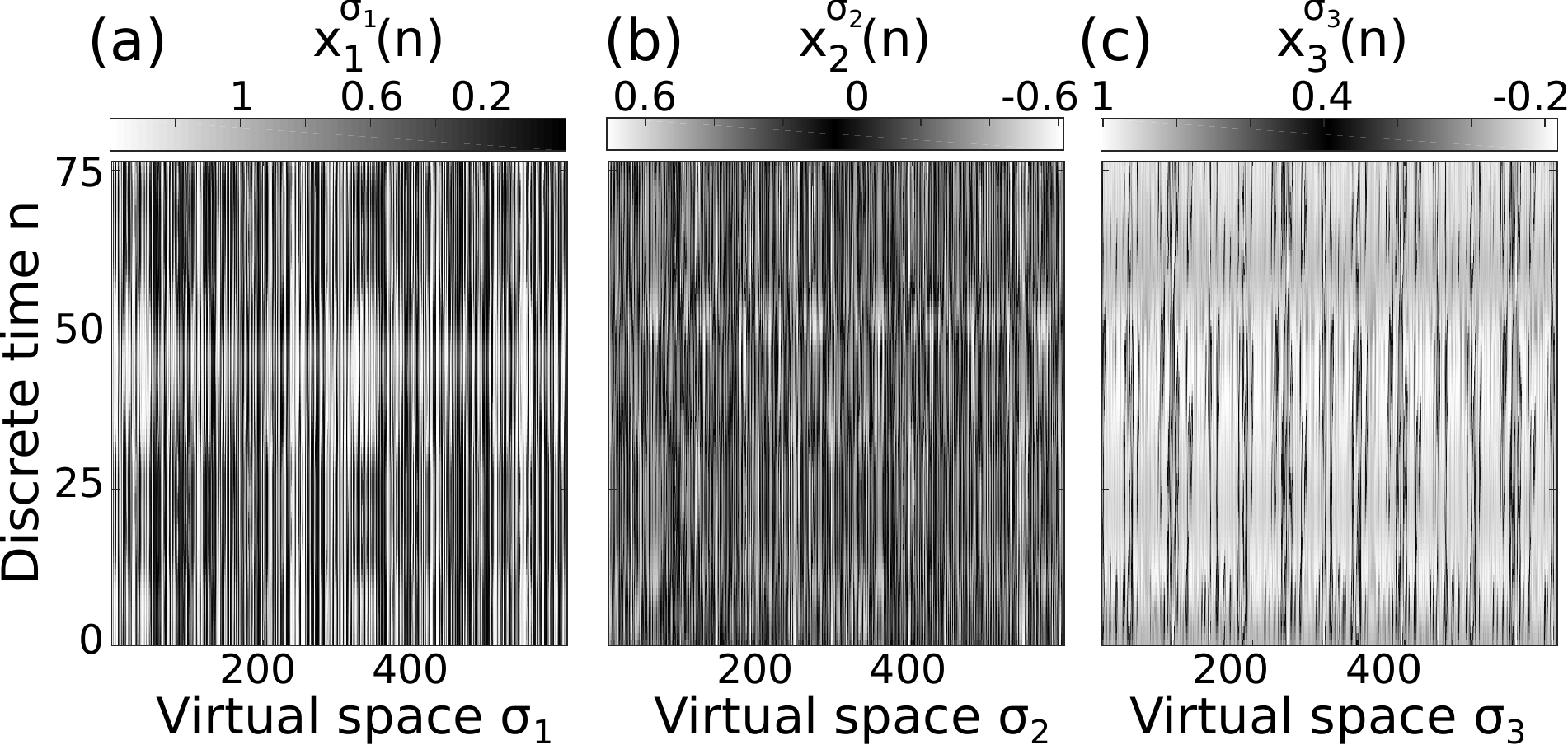}
    \caption{\label{fig:Spatio-temporal}
    Neuron responses ($x^{\sigma_{i}}_{i}(n)$) found in layers $i=1$ (a), $i=2$ (b) and $i=3$ (c), illustrated in a spatio-temporal ($\sigma_{i},n$) representation.
    The spatial frequency along virtual space $\sigma_{i}$ continuously decreases for the higher layers.
    Comparable functionality is implemented in deep convolutional networks or deep reservoirs.}
    \vspace{-0.75cm}
\end{figure} 

According to Eq. \eqref{eq:Arg}, layer $i$ is coupled to layer $i+1$ ($i-1$) according to the fixed connection-weight $w_{i+1,i}$ ($w_{i - 1,i}$), and coupling to $i - 1 = 0$ is unphysical and hence eliminated.
 Therefore, a recurrent layer simply consist of one hardware nonlinearity, one linear delay line and its fixed connections to previous or consecutive layer.
 This has multiple consequences.
 First, inter-layer coupling is instantaneous and constant in time.
 Training of the inter-layer connections, a long-time open question for deep reservoirs \cite{Gallicchio2017} and significant challenge for full hardware integration \cite{Antonik2016}, is therefore avoided.
 Second, such a minimal complexity architecture \cite{Soriano2015} can readily be implemented in hardware \cite{Tanaka2019RecentReview}.
 Finally, it allows establishing a clear mapping from deep TDRs onto deep convolutional neural networks.

The fact that TDR-layers can be termed \textit{convolutional} originates from a nonlinear dynamical node's response to perturbations.
 The state of a nonlinear node in layer $i$ is given by the convolution between its impulse response function $h_{i}(t)$ and its drive $d_{i}(t)$.
 Combined with a normalization of continuous time $t$ by feedback delay $\tau_{Di}$, one can express the dynamical evolution by
\begin{align}
 \frac{t}{\tau_{Di}} &= n + \sigma_{i} / N_{i}, \sigma_{i}\in \{1, N_{i}\}, n\in \{ 1, 2, \dots \}, \label{eq:time} \\
 x_{i}^{\sigma_{i}}(n) &= \int_{-\infty}^{n+\sigma_{i}}h_{i}\left(n+\sigma_{i}-\xi\right) \sin^{2}\left[d_{i}(\xi-1) + b_{i}\right] d\xi \label{eq:convolution} ,
\end{align}
 with $N_{i}$ as the number of neurons in layer $i$, see Fig. \ref{fig:Scheme}.
 Firstly, Eqs. \eqref{eq:time} and \eqref{eq:convolution} map continuous time $t$ onto discrete time $n$ and node $\sigma_{i}$'s position relative to delay time $\tau_{Di}$.
 Details of this temporal embedding technique can be found in \cite{Arecchi1992, Larger2017,Brunner2018}.
 Secondly, expressing the dynamical evolution via the convolution operation shows that a node's impulse response function corresponds to the convolution kernels of a CNN-layer.
 Crucially, coupling created with such a dynamical convolution can directly be translated to the convolution kernel of spatio-temporal networks \cite{Brunner2018,Hart2018}.

\begin{figure*}[ht]
\includegraphics[width=0.85\textwidth]{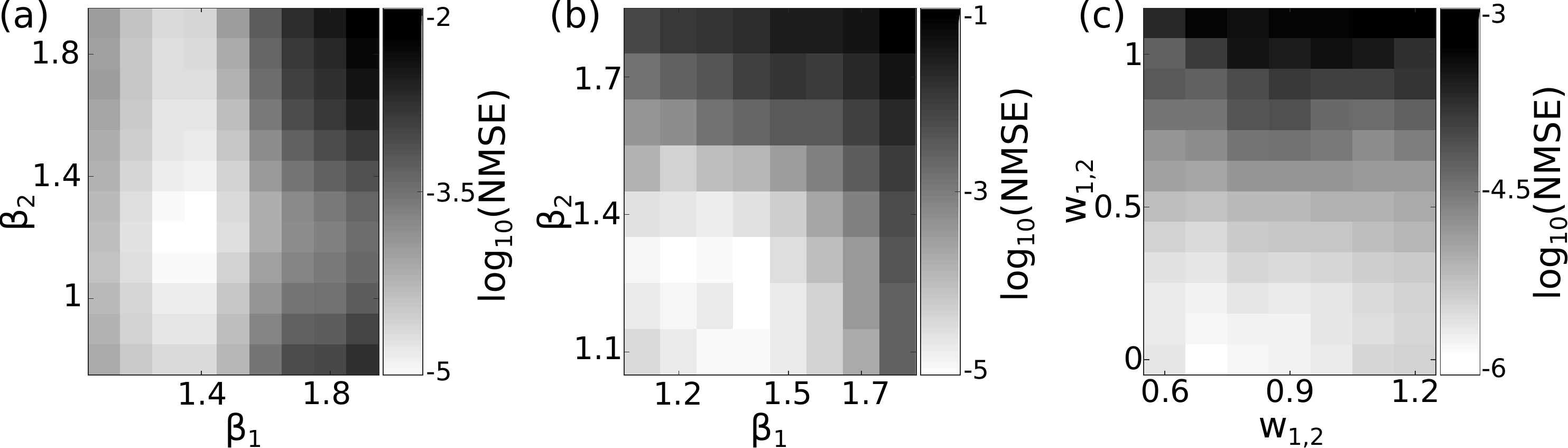}
    \caption{\label{fig:ResultsMG}
    Coupling strongly enhances the network's performance for predicting the chaotic Mackey-Glass timeseries by $\Delta{n}=34$ steps into the future.
    (a) The uncoupled system (both systems receive the input sequence $u(t)$) ($w_{1,2}=w_{2,1}=0$) achieves NMSE=$8.3\cdot10^{-6}$.
    (b) Bidirectional coupling ($w_{1,2}=0.7$, $w_{2,2}=0.6$) results in no improvement (NMSE=$8.8\cdot10^{-6}$).
    (c) The decisively best performing architecture is the unidirectional coupling between the recurrent layers, i.e. feed-forward connections ($w_{1,2}=1.4$, $w_{2,1}=0$): NMSE=$1.3\cdot10^{-6}$.}
    \vspace{-0.5cm}
\end{figure*}
 
The analogy between cascaded TDRs and deep convolutional networks goes further.
 Layers of a CNN commonly feature convolution kernels whose width increases the further back in the cascaded hierarchy a layer is located \cite{Krizhevsky2012ImageNetNetworks}.
 This operation is often associated with generalization: convolution with wider filters reduces the importance of local features in their input, while more general aspects are highlighted.
 The cascaded arrangement of layers in CNNs therefore produces layers which accentuate different input information features.
 In TDRs, increasing the convolution kernel's width corresponds to widening $h_{i}(t)$, see \cite{Supp}.
 Here this is realized by an additional low-frequency cut-off according to timescale $\delta_{i}$ in Eqs. \eqref{eq:DDIE} and \eqref{eq:Int}, and we enforce widening kernels.
 In Fig. \ref{fig:Spatio-temporal} we show the response of a three-layer deep TDR driven by the chaotic Mackey-Glass sequence.
 Each sample corresponds to $\delta t=1$ time-step of the Mackey-Glass system, for which we used the same parameters as in \cite{Jaeger2004}.
 Parameters are $\beta_{2,3}=1.1$, $\tau_{1}=6\cdot{10}^{-3}$, $\tau_{2,3}=7\cdot{10}^{-3}$, $\tau_{Di}=17.85$, $\Phi_0=0.2$, $\delta_{2,3}=0.01$, $w_{1,2}=0.7$, $w_{2,3}=0.8$, $w_{2,1}=w_{3,2}=0$.
 Responses are plotted in spatio-temporal representation \cite{Arecchi1992,Marino2018}, where nodes are arranged along $\sigma_{i}$ and the temporal evolution is along discrete time $n$, with $n$ typically close to a system's delay $\tau_D$ \cite{Brunner2018}.
 As we move into higher layers, from (a) to (c) in Fig. \ref{fig:Spatio-temporal}, dynamics do highlight different spatio-temporal scales.
 Our deep TDR therefore hosts features much like those taken into consideration in the design of CNNs.

Creating a computational result requires to connect the deep TDR to an output via weights adjusted during learning.
 Our readout layer has access to all virtual nodes of all network layers, and the system's output is created according to
\begin{equation}
    y_{j}^{out}(n) = \sum_i^{I} \sum_{\sigma_{i}}^{N_{i}} W_{i,\sigma_{i}, j}^{out} x_{i}^{\sigma_{i}}(n).
\end{equation}
 Here, $j$ is the dimension of the system's output, which depends on the particular task.
 Common methods to obtain $\mathbf{W}^{out}$ are based on linear or ridge regression, and $W^{out}$ is optimized using a representable set of training data \cite{Jaeger2004,Brunner2018}.
 In experimental systems, these methods can be implemented in auxiliary hardware like field-programmable gate arrays \cite{Hermans2016}, or can to a degree be replaced by Boolean learning algorithms \cite{Bueno2018}.
 Recurrent neural networks are primarily relevant for processing temporal information.
 We therefore task the system to predict chaotic sequences $\Delta{n}$ timesteps into the future.
 Training optimizes $W^{out}$ for $y^{out}(n)$ to approximate target $y^{T}(n) = s(n + \Delta{n}), n\in \{1, n^{T} \}$, where $n^{T}=5000$ are the number of samples used for training.
 We quantify the prediction's quality for $n>n^{T}$, hence on testing data not used for training the system, according to the normalized mean square error $NMSE = 1 / n^{T} \sum_{n=1\dots n^{T}} (y^{T}(n) - y^{out}(n))^{2} / (\sigma^{T})^{2}$, where $\sigma^{T}$ is the target-signal's standard deviation.

First, we predict the chaotic Mackey-Glass delay equation, which features a delay of 17 timesteps.
 By predicting ahead twice its delay ($\Delta{n}=34$), the objective is long-term prediction.
 We establish a systematic interpretation by cascading only two TDR layers ($N_{i}$=600) and display the performance dependence on the exhaustively scanned system parameters in Fig. \ref{fig:ResultsMG}.
 We keep $\tau_{1}=0.6\cdot{10}^{-3}$, $\tau_{2}=0.6\cdot{10}^{-3}$, $\tau_{D1,2}=12$, $b_{1,2}=0.2$, $\rho_{1}=8$ and $\delta_{2}=0.01$ constant, with their values mostly based on empirical observations.
 In order to provide a baseline-reference for other topologies, we evaluate uncoupled layers ($w_{1,2}=w_{2,1}=0$) and scan the bifurcation parameter-plane ($\beta_{1}, \beta_{2}$), see Fig. \ref{fig:ResultsMG}(a).
 Importantly, for this test we set $\rho_{2}=\rho_{1}$ and hence couple the BP-layer to the same input as the LP layer.
 We find a clear optimum for $\beta_{1}$, while performance dependence on $\beta_{2}$ is less pronounced.
 The lowest error (NMSE=$8.3\cdot10^{-6}$) is obtained at $\beta_{1}=1.4$ and $\beta_{2}=1.2$.

We now turn to different coupling topologies and disconnect the second layer from the system's input information ($\rho_{2}=0$, $w_{1,2}=0.7$, $w_{2,1}=0.6$).
 Figure \ref{fig:ResultsMG}(b) shows that bidirectional coupling significantly alters the optimal bifurcation parameters and results in a equally pronounced $\beta_{2}$ dependency.
 We obtain NMSE=$8.8\cdot10^{-6}$ at $\beta_{1}=1.4$ and $\beta_{2}=1.2$, and the performance benefit of bidirectional coupling is negligible.
 Continuing with the optimized value of $\beta_{i}$, we focus on the coupling topology by exhaustively scanning $w_{1,2}$ and $w_{2,1}$, see Fig. \ref{fig:ResultsMG}(c).
 The NMSE reveals some performance sensitivity upon the coupling-strength from the first to the second layer.
 The most important finding is, however, that there is a systematic dependency upon $w_{2,1}$: the clear global performance optimum is found for unidirectional coupling with $w_{2,1}=0$.
 The achieved prediction error (NMSE=$1.3\cdot10^{-6}$) is $\sim{3}$ times smaller than for the bidirectional and the uncoupled systems, confirming the benefit of the hierarchical arrangement between consecutive network layers also for TDRs.

To further generalize our finding, we turn to predicting the chaotic Lorenz system.
 The Lorenz system is a three-dimensional set of ordinary differential equations.
 Each sample corresponds to $\delta t=$0.02 time-steps, and we used the same parameters as in \cite{Lu2017}.
 The input information was the Lorenz system's first dimension $x(n)$, and the prediction target was $y^{T}(n) = x(n + 1)$, hence $\Delta{n}=1$.
 Results are listed in Tab. \ref{tab:Lorenz}.
 Prediction performance is again enhanced by the addition of two layers in a unidirectional configuration.
 However, on a first glance the positive benefit appears to be smaller.

Until now prediction only evaluated the system via predicting ahead by distance $\Delta{n}$.
 A more suited approach to determine the capacity of approximating a chaotic system's behavior is based on the so called teacher forcing \cite{Jaeger2004}.
 After training using $\Delta{n}=1$, the system's input becomes its own output, $s(\tilde{n}) = y^{out}(\tilde{n}-1), \tilde{n} = n - n^{T}, n>n^{T}$.
 The TDR becomes an autonomous predictor of the learned system \cite{Jaeger2004}, and the autonomous evolution enables comparison to the original chaotic sequence over long intervals.
 Crucially, this corresponds to predicting until $\tilde{n}$ only relying information of the original signal at $n^{T}$; the prediction autonomously advanced from there.
 This reveals how well the chaotic system as a whole is approximated by the neural network.

\begin{figure}[t]
\includegraphics[width=0.49\textwidth]{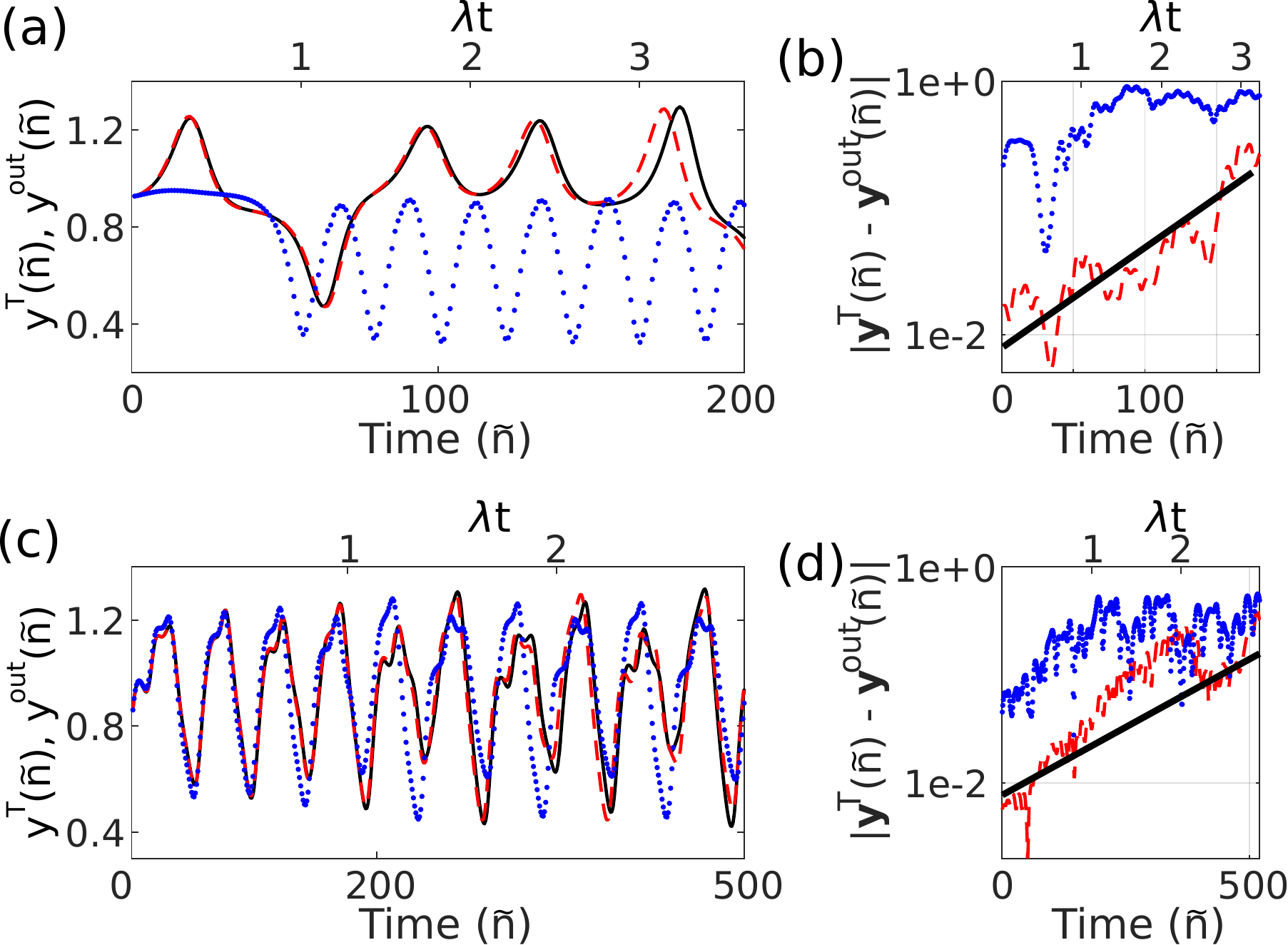}
    \caption{\label{fig:FBFOrcing}
    (Colour online) When connecting the system to its own predicted output at $\tilde{n}=1$, its dynamical evolution becomes autonomous from the original chaotic timeseries.
    The top x-axis is in units of the inverse Lyapunov exponent.
    The long-term prediction performance for predicting the Lorenz (a) and Mackey-Glass system (c) via the connected (not connected) system as red dashed (blue dotted) data.  
    Divergence between the predicted and the original attractors are shown in (b) and (d) for Lorenz and Mackey-Glass, respectively.
    The solid line indicates divergence according to the largest Lyapunov exponent.}
    \vspace{-0.5cm}
\end{figure}

Figure \ref{fig:FBFOrcing} (a) and (c) show autonomous evolution for Lorenz and Mackey-Glass prediction using three cascaded TDR-layers with unidirectional coupling.
 The prediction targets are the black solid data.
 The positive impact of deep (red dashed data) over the single-layer (blue dotted data) TDRs is apparent, and particularly striking when predicting the Lorenz system, see Fig. \ref{fig:FBFOrcing}(a).
 Rather than chaotic excursions along an attractor, the autonomous single layer TDR quickly converges to a dynamical state resembling a limit-cycle and therefore fails to reproduce its target system.
 Only with the three layers coupled in a deep, uni-directional topology the network is capable of an excellent approximate of Lorenz chaos.
 This is also visible from the temporal divergence measured as the Euclidean distances between the Taken's reconstructed attractors of $\mathbf{y}^{out}(\tilde{n})$ and $\mathbf{y}^{T}(\tilde{n})$, see Fig. \ref{fig:FBFOrcing}(b) and (d).
 The solid black lines indicate the divergence according to the maximum Lyapunov exponent (Mackey-Glass: $\lambda_{max}=5.8\cdot10^{-3}$, Lorentz: $\lambda_{max}=0.91$).
 Cascading layer improves prediction by a factor of 20 and 10.5 for Lorenz and Mackey-Glass prediction, respectively.
 The substantial improvement and fundamental importance of the cascaded, 3-layer deep TDR architecture can be further appreciated by inspection of the resulting return maps, see \cite{Supp}.

\begin{table}
\begin{centering}
\begin{tabular}{|c|c|c|}
\hline 
Nodes per layer  & Coupling strength & LZ NMSE\tabularnewline
\hline 
\hline 
1200 lp & -- & $7.6\cdot10^{-7}$\tabularnewline
\hline 
600 lp, 600 bp & $w_{\ensuremath{1,2}}=1.1$ & $5.7\cdot10^{-7}$\tabularnewline
\hline 
400 lp, 400 bp, 400 bp & $w_{\ensuremath{1,2}}=w_{\ensuremath{2,3}}=1.1$ & $2.5\cdot10^{-7}$\tabularnewline
\hline 
\end{tabular}
\par\end{centering}
\caption{\label{tab:Lorenz} Comparison for different architectures with identical total number of neurons $N=1200$.
 Layers are lp=low-pass, bp=band-pass. LZ: Lorenz chaotic time series one step prediction parameters: $\tau_{1}=$ 0.006, $\tau_{2}=\tau_{3}=0.007$, $\delta_{2}=\delta_{3}=0.01$, $\beta_{1}=1.5$, $\beta_{2}=\beta_{3}=1.2$.}
 \vspace{-0.5cm}
\end{table}

We shall finish our investigation by also discussing limitations of our approach.
 The range of possible kernel shapes is limited by physical constraints and have not been optimized during training, through this is possible in principle.
 Also, deep TDRs do no yet reach the accuracy of the original spatio-temporal reservoir \cite{Jaeger2004}.
 Predicting the Mackey-Glass timeseries 84 steps into the future results in NMSE=$10^{-4.4}$ with our deep TDR, while the original reservoir achieves NMSE=$10^{-8.4}$ \cite{Jaeger2004}.
 However, multiple simple additions to the current concept could still significantly improve performance \cite{Martinenghi2012, Grigoryeva2014}.
 Using current high-performance hardware \cite{Jouppi2017}, CNN still run five times slower than TDRs \cite{Larger2017}.
 However CNNs are optimized via back-propagation, which will certainly result in lower errors than deep TDRs.
 If error back propagation can be realized in deep hardware networks remains questionable, while training of our system retains the simplicity and elegance of reservoir computing.

To conclude, we have introduced an elegant scheme for deep convolutional networks in a simple architecture of coupled nonlinear oscillators with delay.
 Information processing conditions conceptually comparable to deep convolutional neural networks with widening convolution kernels are achieved by cascading TDRs with increasingly longer internal timescales.
 Intra- and inter-layer connectivity can be adjusted via the oscillators' time scales, providing a practical control mechanism for hardware realizations.

Applied to both, Mackey-Glass and Lorenz chaos prediction, our concept significantly improves the quality of long-term predictions and proofs essential in the case of Lorenz forecasting.
 Recently, reservoirs have been demonstrated to infer a chaotic oscillator's hidden degrees of freedom \cite{Lu2017} and to predict the evolution of chaotic spatio-temporal systems far into the future \cite{Pathak2018}.
 Temporal structure found in the divergence between prediction and target, such as in Fig. \ref{fig:FBFOrcing}(d), could be addressed via further optimizing timescales $\tau_{i}$ and $\delta_{i}$.

Finally, we would like to point out the large variety of possible hierarchical TDR networks.
 Hybrid systems, where for some or all layers self-feedback is removed, would incorporate feed-forward architectures \cite{Ortin2015}.
 Layers featuring excitable solitons can potentially create long term memory \cite{Romeira2016} and, when combined with the reported LP and BP-layers, physically implement long-short term memory networks \cite{Hochreiter1997}.
 This opens possibilities in new domains like natural language processing and sequence generation.

This work was supported by the EUR EIPHI program (Contract No. ANR-17-EURE-0002), by the BiPhoProc ANR project (No. ANR-14-OHRI-0002-02), by the Volkswagen Foundation NeuroQNet project and the ENERGETIC project of Bourgogne Franche-Comt\'{e}.
 X.P. has received funding from the European Union’s Horizon 2020 research and innovation programe under the Marie Sklodowska-Curie grant agreement No. 713694 (MULTIPLY).

\bibliography{references}

\end{document}